\documentclass[10pt,twocolumn,letterpaper]{article}

\usepackage[pagenumbers]{iccv}      

\usepackage{algorithm}
\usepackage{algpseudocode}
\usepackage{comment}
\usepackage{subcaption}

\definecolor{iccvblue}{rgb}{0.21,0.49,0.74}
\usepackage[pagebackref,breaklinks,colorlinks,allcolors=iccvblue]{hyperref}

\def\paperID{*****} 
\def\confName{ICCV}
\def\confYear{2025}

\title{Extracting Document Relations from Search Corpus by \\Marginalizing over User Queries}

\author{
Yuki Iwamoto \\
{\tt\small iwamoto@sw.cei.uec.ac.jp}
\and
Kaoru Tsunoda \\
{\tt\small tsunoda@uec.ac.jp} 
\and
Ken Kaneiwa \\
{\tt\small kaneiwa@uec.ac.jp} 
\\
\and
The University of Electro-Communications \\
Chofu, Tokyo, Japan
}

\begin{document}
\maketitle
\begin{abstract}
Understanding relationships between documents in large-scale corpora is essential for knowledge discovery and information organization.
However, existing approaches rely heavily on manual annotation or predefined relationship taxonomies.
We propose EDR-MQ (Extracting Document Relations by Marginalizing over User Queries), a novel framework that discovers document relationships through query marginalization.
EDR-MQ is based on the insight that strongly related documents often co-occur in results across diverse user queries, enabling us to estimate joint probabilities between document pairs by marginalizing over a collection of queries.
To enable this query marginalization approach, we develop Multiply Conditioned Retrieval-Augmented Generation (MC-RAG), which employs conditional retrieval where subsequent document retrievals depend on previously retrieved content.
By observing co-occurrence patterns across diverse queries, EDR-MQ estimates joint probabilities between document pairs without requiring labeled training data or predefined taxonomies.
Experimental results show that our query marginalization approach successfully identifies meaningful document relationships, revealing topical clusters, evidence chains, and cross-domain connections that are not apparent through traditional similarity-based methods.
Our query-driven framework offers a practical approach to document organization that adapts to different user perspectives and information needs.
\end{abstract}


\section{Introduction}
\begin{figure}[t]
    \centering
    \includegraphics[width=\linewidth]{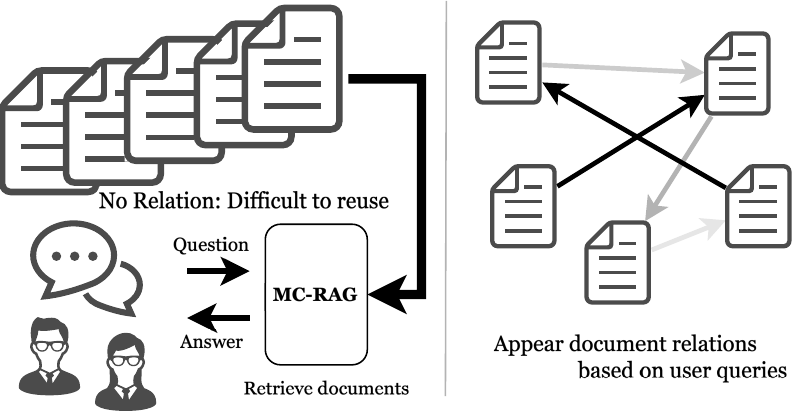}
    \caption{Overview of the EDR-MQ framework for extracting document relationships through user query marginalization}
    \label{fig:overview}
\end{figure}

The exponential growth of digital documents across various domains has created an urgent need for automated methods to understand and organize large-scale document collections~\cite{cui2021bussinessAI}.
While traditional information retrieval systems excel at finding relevant documents for specific queries, they fall short in revealing the underlying relational structure that connects documents within a corpus~\cite{yu2021graph, chen2023towards}.
Understanding these document relationships is crucial for applications ranging from knowledge discovery and semantic search to automated literature review and content recommendation systems~\cite{arslan2024business, cui2021bussinessAI}.

Recent advances in Retrieval-Augmented Generation (RAG) have demonstrated remarkable success in leveraging external knowledge for improved text generation~\cite{lewis2020rag}.
However, while RAG systems effectively retrieve relevant documents for generation, the relationships between documents remain opaque to users.
This limitation makes it difficult for users to verify facts, understand the broader context, or reuse the knowledge for other purposes, as they cannot see how different pieces of information connect to each other within the corpus.

Consider a scenario where teams in large organizations need to navigate vast internal document repositories containing policies, reports, procedures, and project documentation.
While current RAG-based systems can help employees retrieve relevant documents, the relationships between different pieces of organizational knowledge remain hidden, making it difficult for teams to understand how policies connect to procedures, how different projects relate to each other, or how decisions in one department might impact others.
This lack of document relationship visibility hampers collaboration, leads to inconsistent decision-making across teams, and prevents organizations from fully leveraging their accumulated knowledge assets.

In this work, we introduce EDR-MQ (Extracting Document Relations by Marginalizing over User Queries), a novel framework that addresses these limitations by discovering document relationships through query marginalization.
Our key insight is that by observing how documents co-occur across diverse user queries in retrieval systems, we can infer their underlying relationships without requiring explicit supervision or predefined relationship taxonomies.
To enable this approach, we develop a Multiply Conditioned Retrieval-Augmented Generation (MC-RAG) mechanism where subsequent document retrievals are conditioned on previously retrieved content.

EDR-MQ operates on the principle that documents with strong relationships will frequently co-occur when retrieved for semantically related queries.
By marginalizing over a diverse collection of user queries, we can estimate joint probabilities between document pairs, effectively transforming the retrieval process into a relationship discovery mechanism.
This probabilistic framework enables us to construct relationship matrices that capture the strength of associations between documents, facilitating downstream applications such as knowledge graph construction and semantic clustering.

The main contributions of this paper are:
\begin{itemize}
\item We propose EDR-MQ, a novel framework for extracting document relationships by marginalizing over user queries without requiring labeled training data or predefined taxonomies.
\item We develop MC-RAG, a conditional retrieval mechanism that enables the capture of inter-document dependencies by conditioning subsequent retrievals on previously retrieved documents.
\item We demonstrate the effectiveness of EDR-MQ through comprehensive experiments on scientific literature, showing that our method can discover meaningful document relationships and reveal corpus structure.
\end{itemize}

Our experimental results on the SciFact dataset demonstrate that the proposed method successfully identifies meaningful relationships between scientific claims and evidence, outperforming traditional similarity-based approaches in discovering latent document connections.

\section{Related Works}
\subsection{Retrieval Augmented Generation}
Retrieval-Augmented Generation (RAG) has emerged as a powerful paradigm that combines the strengths of parametric and non-parametric approaches to knowledge-intensive natural language processing tasks~\cite{lewis2020rag}.
The core idea of RAG is to augment language models with the ability to retrieve relevant information from external knowledge sources during generation.

The standard RAG architecture consists of two main components: a retriever and a generator.
The retriever, typically implemented using dense passage retrieval methods, identifies relevant documents from a large corpus given a query.
The generator, usually a pre-trained language model such as BART~\cite{lewis2020bart} or T5~\cite{raffel2020t5}, then conditions its generation on both the input query and the retrieved passages.
Formally, RAG computes the probability of generating output $y$ given input $x$ as:
\begin{equation}
    p(y|x) = \sum_{z} p(z|x) p(y|x,z)
\end{equation}
where $z$ represents the retrieved passages, $p(z|x)$ is the retrieval probability, and $p(y|x,z)$ is the generation probability.

Understanding the relationships between retrieved passages in RAG systems remains challenging.
RagViz~\cite{wang2024ragviz} addresses this limitation by visualizing attention patterns between queries and retrieved passages.
However, RagViz focuses on understanding individual query-passage interactions rather than discovering broader relationships between documents across multiple queries.

\subsection{Dense Passage Retrieval and ColBERT}
Dense Passage Retrieval (DPR)~\cite{karpukhin2020dpr} revolutionized information retrieval by replacing traditional sparse retrieval methods (such as BM25) with dense vector representations learned through deep learning.
DPR encodes queries and passages into dense vectors using dual-encoder architectures, typically based on BERT~\cite{devlin2019bert}, and performs retrieval by computing similarity in the embedding space.

While DPR and similar bi-encoder approaches achieve strong retrieval performance, they suffer from the representation bottleneck problem: all information about a passage must be compressed into a single dense vector.
This limitation led to the development of late interaction models such as ColBERT~\cite{khattab2020colbert}.

ColBERT (Contextualized Late Interaction over BERT) addresses the representation bottleneck by postponing the interaction between query and document representations until after encoding.
Instead of producing single vectors, ColBERT generates a sequence of contextualized embeddings for each token in both queries and documents.
The similarity between a query $q$ and document $d$ is computed as:
\begin{equation}
    \text{Score}(q,d) = \sum_{i \in |q|} \max_{j \in |d|} E_q^{(i)} \cdot E_d^{(j)}
\end{equation}
where $E_q^{(i)}$ and $E_d^{(j)}$ are the contextualized embeddings of the $i$-th query token and $j$-th document token, respectively.

\begin{figure*}
    \centering
    \includegraphics[width=0.9\linewidth]{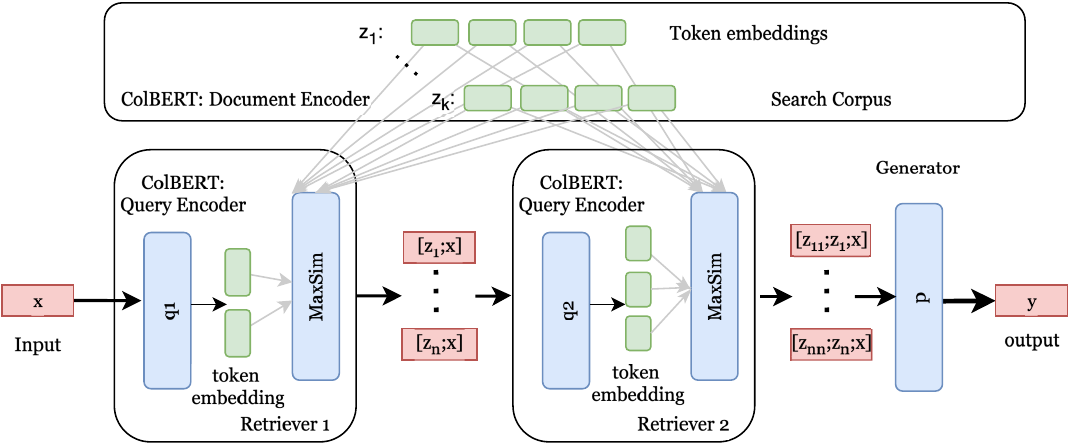}
    \caption{Model Architecture}
    \label{fig: model abst}
\end{figure*}

This late interaction mechanism allows ColBERT to capture fine-grained matching signals between query and document tokens.
ColBERT has demonstrated superior performance across various retrieval benchmarks and has been adopted in numerous applications requiring high-quality document retrieval.

The choice of ColBERT as our retrieval backbone is motivated by its ability to capture nuanced relationships between queries and documents, which is crucial for our conditional retrieval mechanism where the second retriever must understand the relationship between the original query, the first retrieved document, and candidate documents for the second retrieval step.

\textbf{Limitations of Existing Approaches:}
Most existing document relation extraction methods face several limitations:
(1) They require predefined relationship taxonomies or labeled training data;
(2) They operate on static document representations without considering query context;
(3) They cannot adapt to different user perspectives or information needs;

Our approach addresses these limitations by leveraging the query marginalization process in MC-RAG to discover document relationships in an unsupervised manner, without requiring predefined relationship types or labeled training data.
The conditional retrieval mechanism allows us to capture context-dependent relationships that vary based on user queries and information needs.

\section{Question-based Relation Extraction Framework}
In this section, we introduce our proposed framework, which extracts document relations by marginalizing over user queries.
Our framework extracts document relations using MC-RAG's retrievers.
MC-RAG is a RAG that has multiple retrievers to retrieve evidence sentences and then generates answer text. 
Firstly, we introduce MC-RAG in the following section.
Next, we describe how to extract document relations using MC-RAG by marginalizing over user queries.
\subsection{Multiply Conditioned RAG}
MC-RAG is a RAG model that takes a user input $x$ to retrieve multiple text documents $z_{i}, \ldots, z_{k}$,
which are then used as additional context for generating the answer sentence $y$.
Without loss of generality, we describe our model with $k=2$.
MC-RAG has $k$ retrievers ($k=2$): $p_{\eta_{1}}(z_{i}|x)$ and $p_{\eta_{2}}(z_{j}|z_{i}, x)$, which return distributions over text passages given input $x$ and $(z_{i}, x)$, respectively.

We employ ColBERT as the encoder for both retrievers, which computes fine-grained similarity scores between query and document representations.
ColBERT encodes queries and documents into contextualized embeddings at the token level, enabling efficient and accurate retrieval through late interaction mechanisms.
The similarity computation follows ColBERT's approach, where the similarity between query $q$ and document $d$ is computed as:
\begin{equation}
    \text{sim}(q, d) = \sum_{i \in |q|} \max_{j \in |d|} E_q^i \cdot E_d^j
\end{equation}
where $E_q^i$ and $E_d^j$ are the contextualized embeddings of the $i$-th query token and $j$-th document token, respectively.

The key innovation of MC-RAG lies in its conditional retrieval mechanism $p_{\eta_{2}}(z_{j}|z_{i}, x)$.
Unlike standard RAG models that perform independent retrieval, the second retriever conditions its search on both the user input $x$ and the previously retrieved passage $z_i$.
This conditional retrieval is achieved by concatenating the embeddings of $x$ and $z_i$ as the query representation for the second retriever.
Specifically, the second retriever computes similarities using the combined context $[z_i; x]$ where $[;]$ denotes concatenation.

This conditional mechanism is crucial for enabling the computation of joint probabilities $p(z_i, z_j)$ as described in the subsequent section.
A single retriever cannot capture such dependencies between retrieved passages, as it operates independently for each retrieval step.
The conditional structure allows MC-RAG to model relationships between documents and extract structured information from the corpus by marginalizing over user queries.

Similar to standard RAG models, MC-RAG has a generator $p_{\theta}(y|z_{i}, z_{j}, x)$ that produces the output $y$ conditioned on both the input $x$ and the retrieved passages $z_{i}$ and $z_{j}$.
The final output distribution is obtained by marginalizing over all possible combinations of retrieved passages:
\begin{equation}
    p(y|x) = \sum_{z_{i}} \sum_{z_{j}} p_{\eta_{1}}(z_{i}|x)
    p_{\eta_{2}}(z_{j}|z_{i},x) p_{\theta}(y|x, z_{i}, z_{j})
\end{equation}

\subsection{Extracting Document Relations by Marginalizing over User Queries}
In this section, we describe how to extract document relations by marginalizing over user queries.
We calculate the joint probability of the retrieved passages with the following equation:
\begin{equation}
    p(z_{i}, z_{j}) = \sum_{x} p_{\eta_{1}}(z_{i}|x)
    p_{\eta_{2}}(z_{j}|z_{i},x) p(x)
\end{equation}
Therefore, our method leverages the diversity of user queries to estimate the underlying relationships between documents in the corpus.
The more diverse and comprehensive the set of user queries, the more accurate the joint probability estimation becomes, as it captures different aspects and contexts in which documents $z_i$ and $z_j$ are relevant together.

The key insight is that by marginalizing over a large collection of user queries, we can discover latent relationships between documents that may not be apparent from individual queries alone.
Each query $x$ provides a different perspective on how documents relate to each other, and the aggregation across multiple queries reveals stable patterns of document co-occurrence and dependency.
This approach effectively transforms the retrieval process into a structure discovery mechanism, where the conditional retrieval patterns learned by MC-RAG expose the underlying semantic relationships in the corpus.

In practice, we collect a diverse set of queries $\mathcal{X} = \{x_1, x_2, \ldots, x_N\}$ and approximate the marginal distribution as:
\begin{equation}
    p(z_{i}, z_{j}) \approx \frac{1}{N} \sum_{n=1}^{N} p_{\eta_{1}}(z_{i}|x_n) p_{\eta_{2}}(z_{j}|z_{i},x_n)
\end{equation}
where we assume a uniform prior $p(x) = \frac{1}{N}$ over the query set.
The resulting joint probabilities $p(z_i, z_j)$ form a relationship matrix that captures the strength of associations between different document pairs in the corpus.

\subsubsection{Algorithm for Document Relation Extraction}
We now present the detailed algorithm for extracting document relations using MC-RAG.
\cref{alg:doc_relation} outlines the complete process from query collection to relationship matrix construction.

\begin{algorithm}[ht]
\caption{Document Relation Extraction via Query Marginalization}
\label{alg:doc_relation}
\begin{algorithmic}[1]
\Require Document corpus $\mathcal{D} = \{d_1, d_2, \ldots, d_M\}$, Query set $\mathcal{X} = \{x_1, x_2, \ldots, x_N\}$, Top-k retrieval parameter $k$
\Ensure Relationship matrix $R \in \mathbb{R}^{M \times M}$
\State Initialize relationship matrix $R$ with zeros
\State Precompute document embeddings using ColBERT encoder
\For{each query $x_n \in \mathcal{X}$}
    \State $\mathcal{Z}_1 \leftarrow$ Retrieve top-$k$ documents using first retriever $p_{\eta_1}(z_i|x_n)$
    \For{each retrieved document $z_i \in \mathcal{Z}_1$}
        \State Construct conditional query $q_{cond} = [z_i; x_n]$
        \State $\mathcal{Z}_2 \leftarrow$ Retrieve top-$k$ documents using second retriever $p_{\eta_2}(z_j|z_i, x_n)$
        \For{each retrieved document $z_j \in \mathcal{Z}_2$}
            \State $p_1 \leftarrow p_{\eta_1}(z_i|x_n)$ \Comment{First retriever probability}
            \State $p_2 \leftarrow p_{\eta_2}(z_j|z_i, x_n)$ \Comment{Second retriever probability}
            \State $R[i,j] \leftarrow R[i,j] + p_1 \cdot p_2$ \Comment{Accumulate joint probability}
        \EndFor
    \EndFor
\EndFor
\State \Return Relationship matrix $R$
\end{algorithmic}
\end{algorithm}

The algorithm operates in three main phases:

\textbf{Preprocessing Phase (Lines 1-2):} We initialize the relationship matrix and precompute document embeddings using ColBERT to enable efficient similarity computation during retrieval.

\textbf{Query Processing Phase (Lines 3-12):} For each query in the collection, we perform the two-stage retrieval process. The first retriever identifies relevant documents based on the query alone, while the second retriever finds documents that are relevant given both the query and the previously retrieved document. The joint probabilities are accumulated in the relationship matrix, effectively marginalizing over the query distribution.


\begin{figure*}[t]
    \centering
    \begin{subfigure}[b]{0.48\linewidth}
        \centering
        \includegraphics[width=\linewidth]{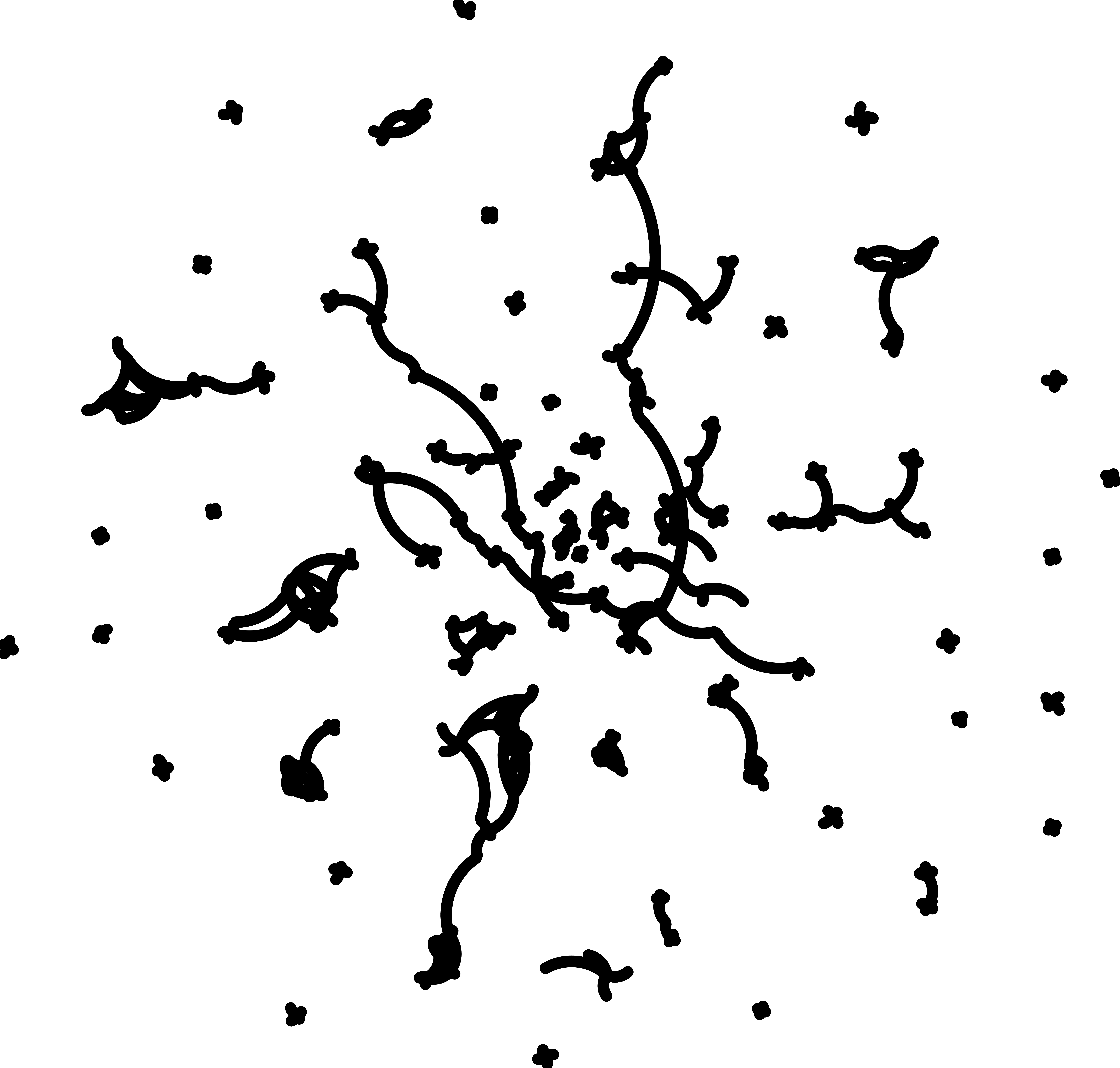}
        \caption{Document relationships with 30 queries}
        \label{fig:few_queries}
    \end{subfigure}
    \hfill
    \begin{subfigure}[b]{0.48\linewidth}
        \centering
        \includegraphics[width=\linewidth]{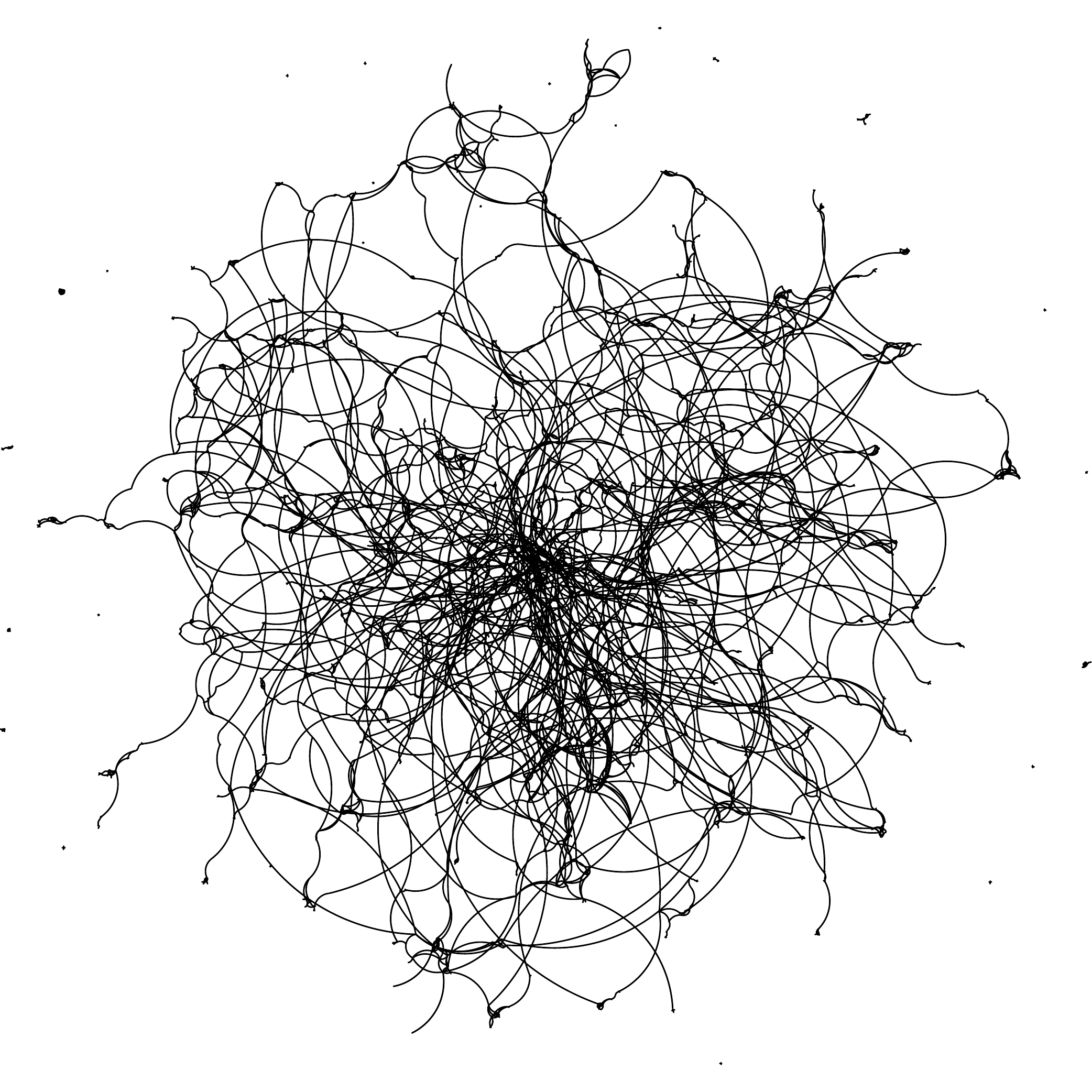}
        \caption{Document relationships with 300 queries}
        \label{fig:many_queries}
    \end{subfigure}
    \caption{Effect of query diversity on document relationship extraction. (a) With only 30 queries (1/10 of the full set), the method creates numerous small, isolated clusters with limited inter-cluster connections. (b) With 300 queries, the relationships become more comprehensive, connecting previously isolated clusters and revealing the overall corpus structure through richer inter-document connections.}
    \label{fig:query_effect}
\end{figure*}

\begin{figure*}
    \centering
    \includegraphics[width=.8\linewidth]{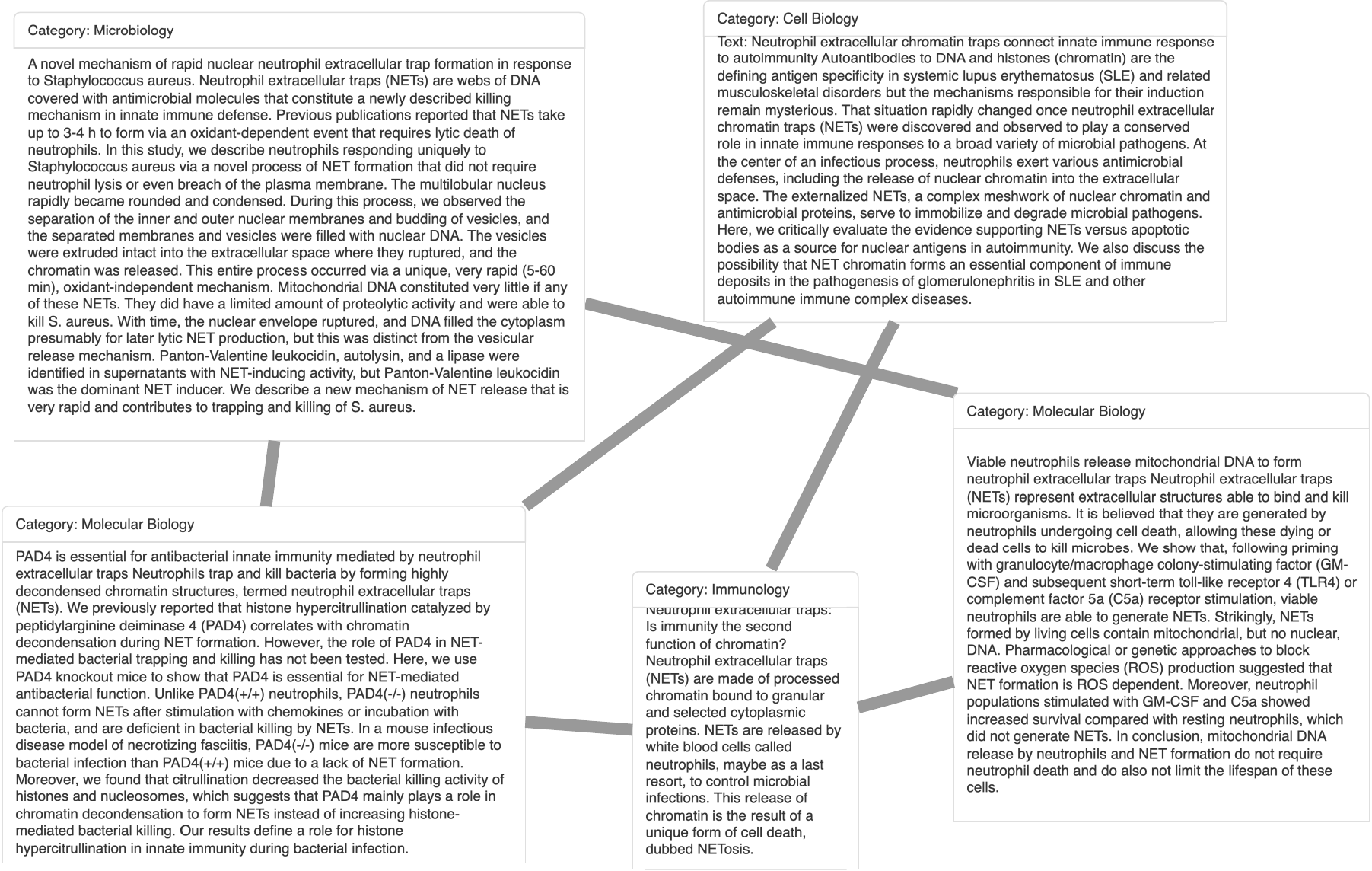}
    \caption{Concrete example of extracted document relationships across different scientific categories. The figure shows how our MC-RAG framework discovers meaningful connections between documents from different domains: Microbiology, Cell Biology, Molecular Biology (two instances), and Immunology. Despite belonging to different categories, the documents are connected through shared concepts such as neutrophil extracellular traps (NETs), bacterial immunity mechanisms, and cellular processes. This demonstrates the method's ability to identify cross-domain relationships that would not be apparent through traditional category-based organization.}
    \label{fig:cross_domain_relations}
\end{figure*}

\begin{figure}
    \centering
    \includegraphics[width=\linewidth]{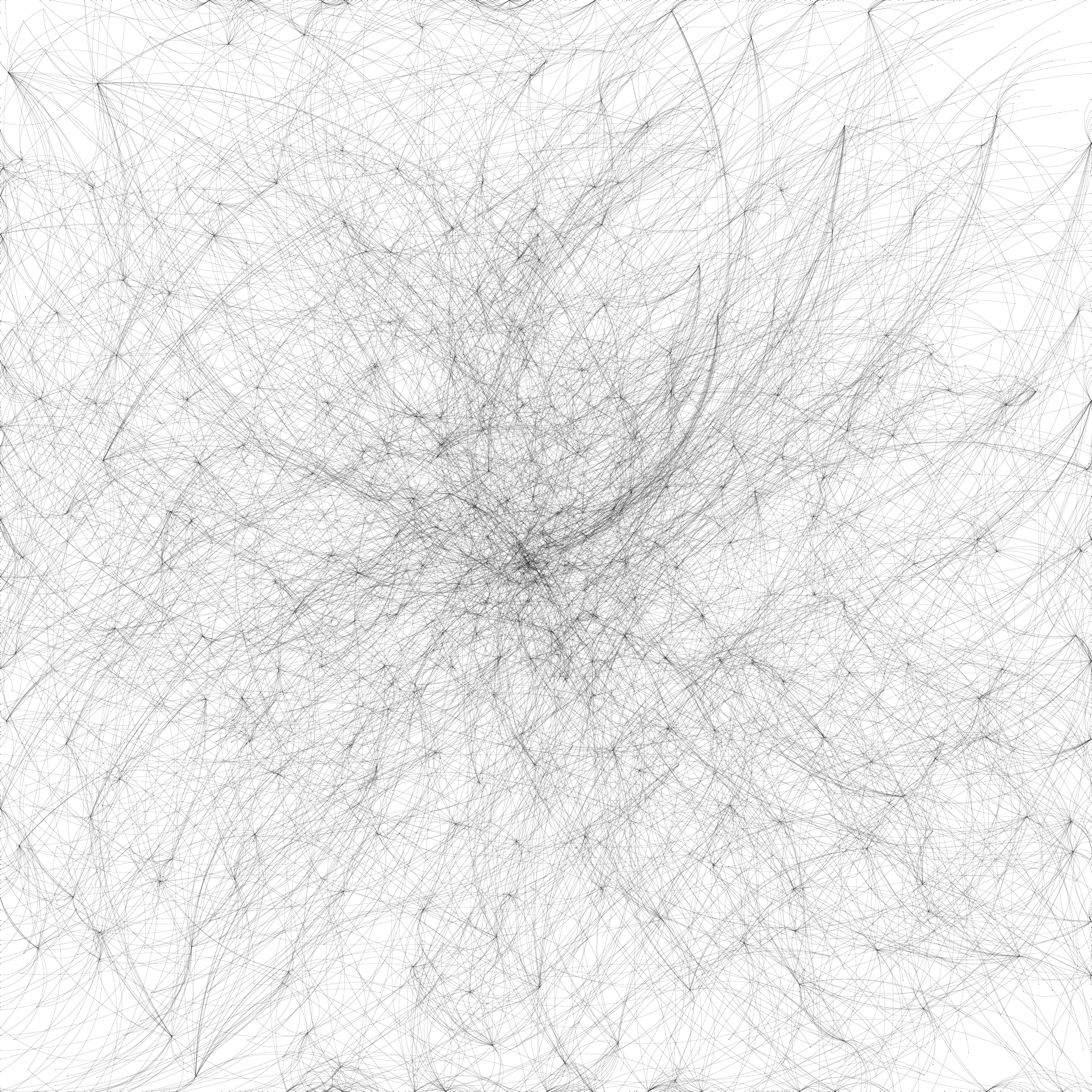}
    \caption{Document relationship network using TF-IDF similarity with top-25 connections per document. The network exhibits excessive complexity with dense interconnections that obscure meaningful patterns, making it difficult to interpret document relationships and identify coherent clusters.}
    \label{fig:tfidf_baseline}
\end{figure}

The computational complexity of this algorithm is $O(N \cdot k^2 \cdot C)$, where $N$ is the number of queries, $k$ is the number of retrieved documents per stage, and $C$ is the cost of similarity computation in ColBERT. The space complexity is $O(M^2)$ for storing the relationship matrix, where $M$ is the total number of documents in the corpus.

\section{Experiments}
In this section, we demonstrate the effectiveness of our proposed EDR-MQ framework through experiments.
We focus on qualitative analyses and visualization results that showcase the method's ability to extract meaningful document relationships and reveal corpus structure as the diversity of user queries increases.

\subsection{Experimental Setup}
We evaluate our EDR-MQ framework on the SciFact dataset~\cite{wadden2020scifact}, which contains 1,409 scientific claims paired with 5,183 abstracts.
For our experiments, we treat each claim as a user query, and paper abstracts as a corpus.

For visualization, we employed Gephi with the ForceAtlas 2 algorithm for graph layout, using the parameters summarized in Table~\ref{tab:forceatlas2_params}.

\subsection{Effect of Query Diversity on Relationship Discovery}
A fundamental hypothesis of our approach is that increasing the diversity and number of user queries leads to more comprehensive document relationship extraction.
To validate this hypothesis, we conducted experiments using different numbers of queries on the SciFact dataset and analyzed the resulting relationship networks.

Figure~\ref{fig:query_effect} illustrates the effect of query diversity on relationship discovery.
When using only 30 queries (Figure~\ref{fig:few_queries}), which represents 1/10 of our full query set, the extracted relationship network exhibits numerous small, isolated clusters.
While the method successfully identifies local relationships within these clusters, the limited query diversity fails to reveal connections between different clusters, resulting in a fragmented understanding of the corpus structure.

In contrast, when the full set of 300 queries is employed (Figure~\ref{fig:many_queries}), the relationship network becomes significantly more connected and comprehensive.
The previously isolated clusters are now linked through bridge documents that connect different parts of the corpus.
This demonstrates that our query marginalization approach uncovers more latent relationships as the number of queries increases, revealing connections that are not apparent when using limited queries.

The key insights from this experiment are:
\begin{itemize}
\item \textbf{Query Diversity is Critical}: More diverse queries enable the discovery of relationships across different parts of the corpus, preventing the formation of isolated clusters.
\item \textbf{Bridge Document Discovery}: With sufficient query diversity, documents that serve as bridges between different clusters become apparent, revealing the interconnected nature of the corpus.
\item \textbf{Scalable Relationship Extraction}: The method gracefully scales with query diversity, progressively revealing corpus structure without requiring additional supervision.
\end{itemize}

This experiment supports our theoretical framework where joint probabilities $p(z_i, z_j)$ become more accurate as we marginalize over a larger and more diverse set of user queries, leading to better estimation of document relationships.

\begin{figure*}[t]
    \centering
    \begin{subfigure}[b]{0.48\linewidth}
        \centering
        \includegraphics[width=\linewidth,bb=0 0 1000 800]{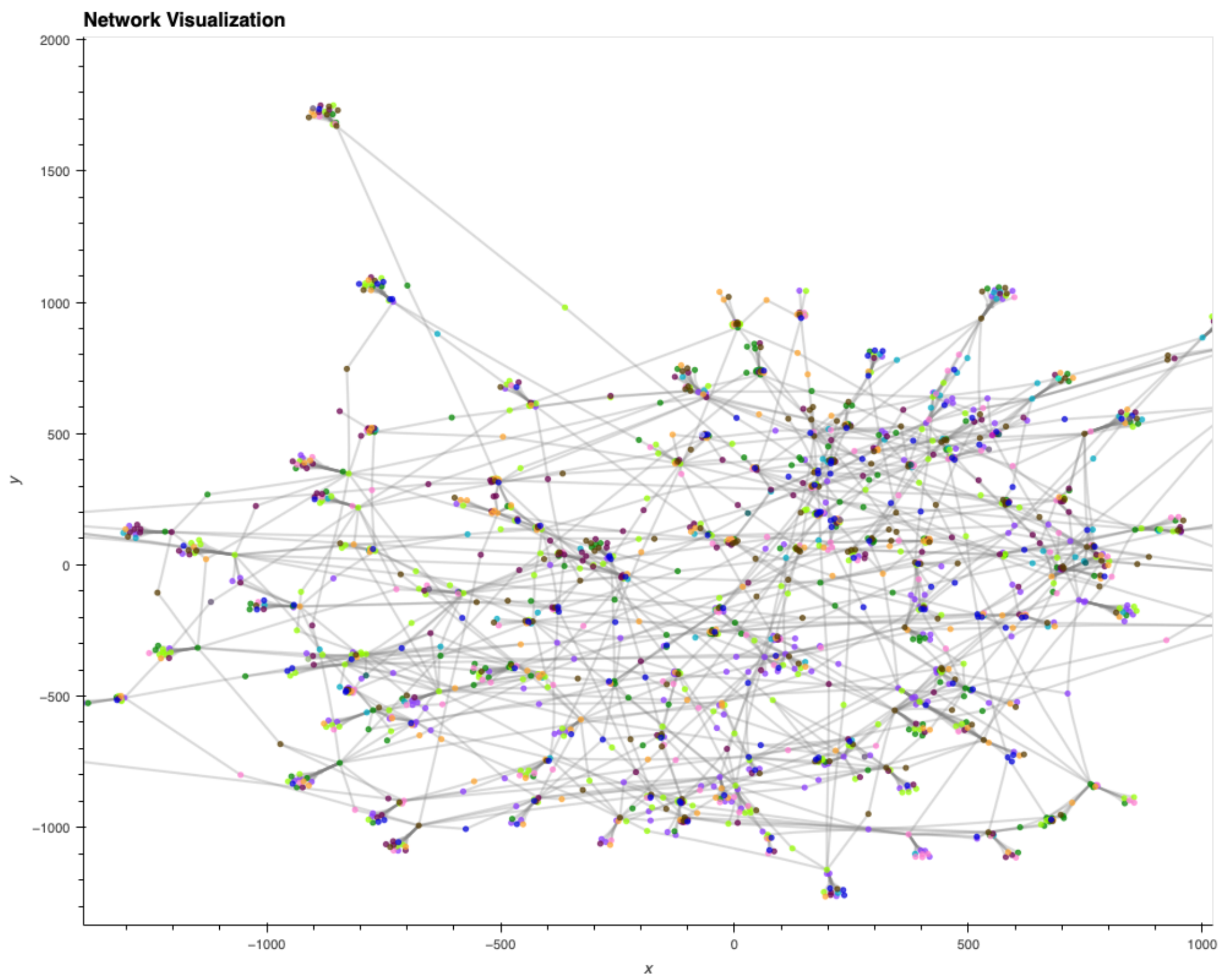}
        \caption{TF-IDF similarity network with edge bundling}
        \label{fig:tf-idf_edge-bundling}
    \end{subfigure}
    \hfill
    \begin{subfigure}[b]{0.48\linewidth}
        \centering
        \includegraphics[width=\linewidth,bb=0 0 1000 800]{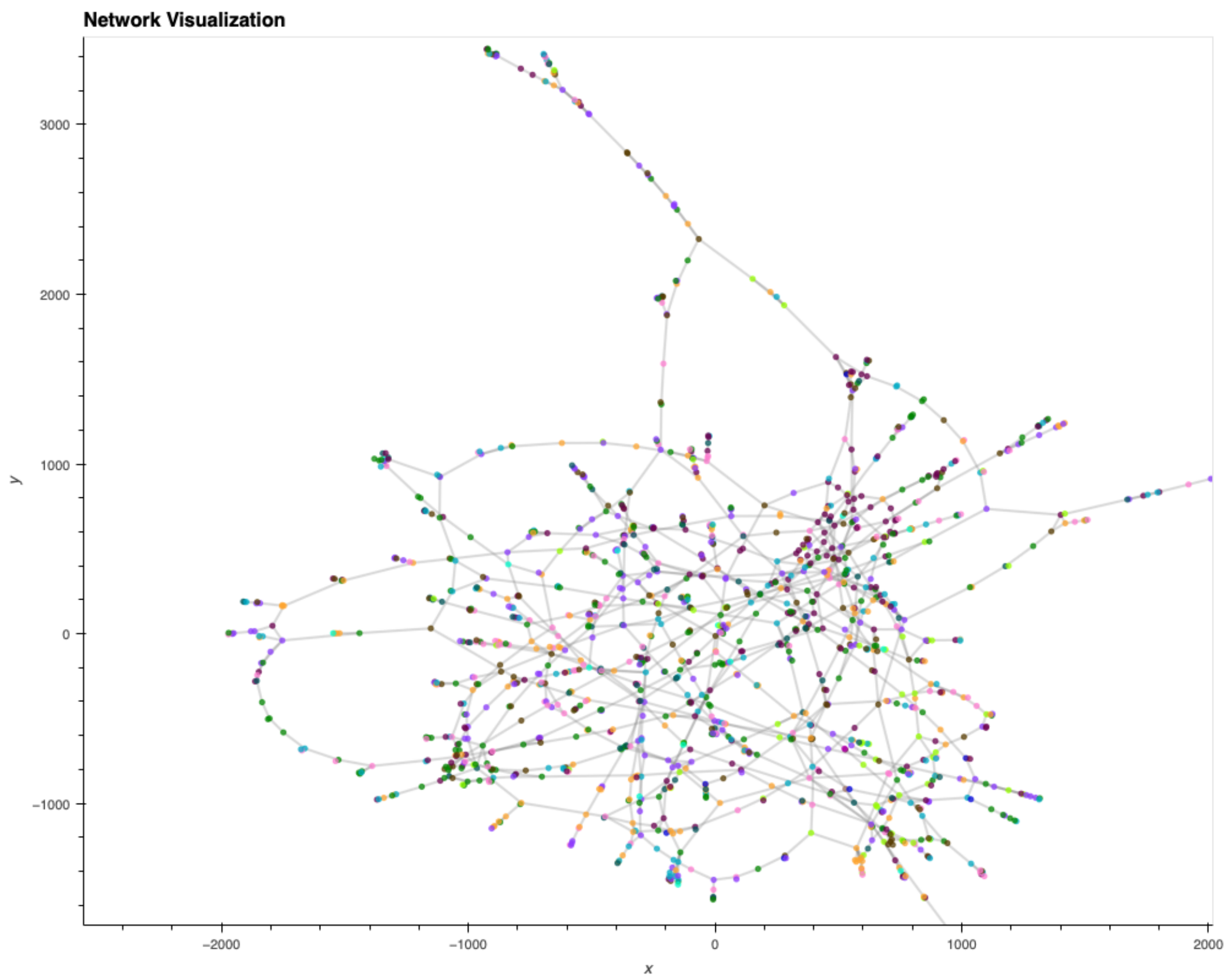}
        \caption{EDR-MQ relationship network with edge bundling}
        \label{fig:edr-mq_edge-bundling}
    \end{subfigure}
    
    \caption{Comparison of document relationship networks using edge bundling visualization. Documents are color-coded by category and positioned using ForceAtlas 2 layout. (a) TF-IDF similarity network shows scattered edge patterns with poor bundling, making overall trends difficult to interpret. (b) EDR-MQ network exhibits clear edge bundling patterns that reveal coherent relationship structures and facilitate interpretation.}
    \label{fig:edge-bundling}
\end{figure*}

\subsection{Network Interpretability and Visualization Analysis}
To demonstrate the advantages of our EDR-MQ framework, we compare it with traditional document similarity approaches through comprehensive network analysis and visualization.

Figure~\ref{fig:tfidf_baseline} shows the document relationship network constructed using TF-IDF similarity, where each document is connected to its top-25 most similar documents based on term frequency-inverse document frequency scores.
This top-25 setting corresponds to the same number of document pairs that our EDR-MQ method considers when performing 5 retrievals in the first stage followed by 5 retrievals in the second stage (5×5=25).

The TF-IDF-based network exhibits several problematic characteristics that highlight the limitations of traditional similarity measures.
The network is densely interconnected, creating a tangled web that obscures meaningful relationship patterns.
This excessive complexity makes it extremely difficult to identify coherent document clusters or understand the semantic basis of connections.
Moreover, most documents have similar numbers of connections, failing to distinguish between central hub documents and peripheral nodes.
Unlike our EDR-MQ results, the TF-IDF network lacks a clear community structure or hierarchical organization, presenting an absence of interpretable structure.

To further analyze the structural differences between traditional similarity measures and our EDR-MQ approach, we applied edge bundling visualization to both networks.
Edge bundling~\cite{zhou2013edge} is a graph visualization technique that groups similar edges together into bundles, reducing visual clutter and revealing high-level connectivity patterns in complex networks.
This technique is particularly useful for understanding the overall flow and structure of document relationships.

Figure~\ref{fig:edge-bundling} presents a compelling comparison between TF-IDF similarity networks and our EDR-MQ approach using edge bundling visualization.
Documents are color-coded by their categorical labels and positioned using the ForceAtlas 2 layout algorithm.
The edge bundling technique groups related connections together, making it easier to identify major relationship pathways and structural patterns.

The TF-IDF-based network (Figure~\ref{fig:tf-idf_edge-bundling}) exhibits scattered edge patterns with poor bundling characteristics.
The edges appear dispersed across the visualization space with little coherent grouping, making it difficult to discern meaningful relationship trends or identify major connection pathways between document categories.
This scattered pattern reflects the lexical nature of TF-IDF similarity, which often produces connections based on surface-level term overlap rather than deeper semantic relationships.

In contrast, the EDR-MQ network (Figure~\ref{fig:edr-mq_edge-bundling}) demonstrates clear and well-defined edge bundling patterns.
The relationships form coherent bundles that connect related document clusters, creating interpretable pathways that reflect meaningful semantic connections.
These bundled edges reveal the underlying structure of how different document categories relate to each other through query-driven discovery, making the overall network organization much more comprehensible.

This comprehensive comparison demonstrates that our EDR-MQ approach produces more structured and interpretable document relationships compared to traditional similarity measures, facilitating better understanding of corpus structure and enabling more effective navigation of document collections.

\subsection{Cross-Domain Relationship Discovery}
An important characteristic of our approach is its ability to discover relationships between documents that span different categorical boundaries based on the diversity of user queries.

Figure~\ref{fig:cross_domain_relations} presents a concrete example of how our EDR-MQ framework extracts meaningful relationships between scientific documents from diverse categories.
The visualization shows five documents from four different biological domains:
\begin{itemize}
\item \textbf{Microbiology}: Focuses on neutrophil extracellular trap formation in response to \emph{Staphylococcus aureus}
\item \textbf{Cell Biology}: Discusses neutrophil extracellular chromatin traps and their connection to autoimmune responses
\item \textbf{Molecular Biology} (Document 1): Examines PAD4's role in antibacterial immunity and neutrophil extracellular traps
\item \textbf{Molecular Biology} (Document 2): Investigates viable neutrophils and mitochondrial DNA release in NET formation
\item \textbf{Immunology}: Studies neutrophil extracellular traps and chromatin granule function
\end{itemize}

Despite being classified into different categories, our method successfully identifies the underlying conceptual connections between these documents.
The extracted relationships reveal a coherent research narrative centered around neutrophil extracellular traps (NETs), bacterial immunity mechanisms, and cellular processes involved in immune defense.

This cross-domain relationship discovery capability is further validated by the edge bundling analysis shown in Figure~\ref{fig:edge-bundling}.
The clear bundling patterns in our EDR-MQ network (Figure~\ref{fig:edr-mq_edge-bundling}) demonstrate how documents from different categories are connected through meaningful semantic pathways, whereas the TF-IDF approach fails to reveal such coherent cross-domain connections.
The superior network organization in our approach enables the identification of interdisciplinary relationships that would be missed by traditional category-based organization methods.

This example demonstrates that our query marginalization approach can transcend traditional categorical boundaries to reveal the true conceptual structure underlying scientific literature.
Such cross-domain relationship discovery is particularly valuable for interdisciplinary research, where important insights often emerge from connections between seemingly disparate fields.

\begin{table}[t]
\centering
\caption{ForceAtlas 2 Layout Parameters}
\label{tab:forceatlas2_params}
\begin{tabular}{lc}
\hline
\textbf{Parameter} & \textbf{Value} \\
\hline
Tolerance (speed) & 1.0 \\
Approximate Repulsion & Enabled \\
Approximation & 1.2 \\
Scaling & 2.0 \\
Stronger Gravity & Disabled \\
Gravity & 1.0 \\
Dissuade Hubs & Enabled \\
LinLog mode & Enabled \\
Prevent Overlap & Enabled \\
Edge Weight Influence & 1.0 \\
Normalize edge weights & Enabled \\
Inverted edge weights & Enabled \\
\hline
\end{tabular}
\end{table}

\section{Conclusion}
We introduced EDR-MQ, a novel framework for extracting document relationships through query marginalization using Multiply Conditioned RAG (MC-RAG).
Our approach discovers latent relationships without requiring manual annotation or predefined taxonomies by marginalizing over diverse user queries.

The key innovation lies in the conditional retrieval mechanism, where subsequent retrievals depend on previously retrieved content.
This enables the construction of relationship matrices that reveal corpus structure through query-driven co-occurrence patterns.

Experimental results on the SciFact dataset demonstrate that query diversity is critical for comprehensive relationship extraction, and our method successfully identifies cross-domain relationships that transcend traditional categorical boundaries.
The unsupervised nature of our approach makes it particularly valuable for domains where manual relationship annotation is prohibitively expensive.

In the future, we plan to explore the following directions:
First, end-to-end training of the entire MC-RAG framework could potentially improve the quality of extracted relationships by jointly optimizing the retrieval and relationship extraction objectives.
Second, incorporating advanced edge bundling techniques that provide localized zoom views~\cite{akiyama2024focus+} could enable better understanding of both local and global document relationships.
Finally, comprehensive evaluation across diverse datasets and comparison with additional baseline methods would strengthen the validation of our approach's generalizability.

{
    \small
    \bibliographystyle{ieeenat_fullname}
    \bibliography{main}
}
\end{document}